# Knowledge Management Skills for 21st Century Library Professionals in India: A Study

Subaveerapandiyan A  
*Regional Institute of Education Mysore*, subaveerapandiyan@gmail.com

Dharmavarapu Sindhu  
*National Institute of Technology Puducherry*, sindhuasstlibrarian.nitpy@gmail.com

# Knowledge Management Skills for 21st Century Library Professionals in India: A Study

**Subaveerapandiyan A**
Professional Assistant
Regional Institute of Education Mysore, India
E-mail: subaveerapandiyan@gmail.com
**Dharmavarapu Sindhu**
Assistant Librarian
Central Library
National Institute of Technology Puducherry
E-Mail: sindhuasstlibrarian.nitpy@gmail.com

**Abstract**

*In any kind of educational institute and organization, libraries are playing a crucial role. For the development of library services, skillful library professionals are indispensable. Without knowledge management skills, no one can provide essential services to the users. "Library is a growing organism" by S.R. Ranganathan (1931), based on his law, as library professionals have to adopt futuristic skills for better user services. The study investigates knowledge management skills and the chief strengths of library and information science professionals in India. For this study, data collected from the Institute of National Importance and Central Universities around India. The principal aim of this study is to understand the current scenario and data literacy skills of library professionals. The study found that 98.7% of respondents qualified for library and information science postgraduate and above. They highlighted the very important skills respondents are digital literacy skills (65.9%), soft skills (70%), organization skills (67.7%), leadership skills (70%), and English language skills (52.4%).*

1. Introduction

Libraries have been and will be indispensable in the dissemination of knowledge and its fruit to all walks of human society. With the advancement of science and technology and information and communication technology, it became inevitable for libraries all over the world to switch to sophisticated library infrastructure and novel modes of dispensing knowledge viz., virtual and e-resources, etc.; to serve the increasing third for knowledge among ever-increasing researchers, student community, and readers. This orientation of modern libraries towards tech-savvy infrastructure and methodologies is interesting for library professionals of this century to adapt to emerging trends of library technologies and etiquettes. All the advancement of modern libraries must have been possible with the efforts of library professionals directly or indirectly. As it is well said that "Necessity is the mother of invention", only working library professionals know the underlying necessity for change libraries methods of knowledge dispensation and information required, which eventually lead to evaluation of libraries present state of art infrastructure and technologies like RFID technologies, cloud computing, virtual reality, artificial intelligence and so on in today's libraries world. Library professionals should well be versed with basic operational and troubleshooting procedures of contemporary and upcoming best practices to attract readers who may bacon of light with their future endeavor's and achievements to better both tangible and intangible life on this planet Earth. Our study took experts of skills and competencies possessed by Library Professionals from various Academic Institutions of National importance in the Country. Since mere possession of knowledge can serve no purpose until they put it is pertinent to emphasize the need to employ the skills and competencies of advanced knowledge in the ways they ought to be. The gap between knowledge possessed and that of being possessed is to be filled with concrete policy decisions which may include intensive training of library professionals in their respective areas of importance and conducting of Parliament of Librarians of eminent institutions and decentralized summits of librarians of universities and institutions affiliated to them analogous to the Legislative and executive set up in the country for exchanges of knowledge, to discuss various challenges and to mimic the best practices of the library arena.

## 2. Related Studies

Ranganathan (1931) emphasizes "save the time of the reader". Time management is most prominent in day-to-day life. They manage their time perfectly so they can achieve success in their career. In the library, professionals can save the users' time such as a timely reply for their queries, providing a current awareness service, reprography service, reference service and procurement of books and resources on time and after procurement completes the book process as much as earlier and delivers the book to the users. Time management is stress-free to achieve the vision and mission of the library. Technology supports libraries for managing time such as RFID, Barcode, QR code, Kiosks, Book lifts, etc.

Ugwu & Ezema (2010) surveyed the 47th Annual Conference/AGM of the Nigerian Library Association. They have distributed a questionnaire to 67 people based on their responses; they have identified the crucial competencies: knowledge management skills, cultural skills, leadership skills, strategic skills, and restructuring skills. These are the essential skills for librarians.

ALA's Core Competence of Librarianship (2010) divides into eight categories: i) Foundations of the Profession (Effective communication skill, analytical and problem-solving skill) ii) Information Resources (electronic resources, preservation and conservation skill) iii) Organization of Recorded Knowledge and Information (cataloguing, metadata, indexing, and classification skill) iv) Technological Knowledge and Skills (emerging technologies, innovations, database and programming skill) v) Reference and User Services (Information retrieval and Information literacy skill) vi) Research (quantitative and qualitative research skill) vii) Continuing Education and Lifelong Learning (teaching and creativity skill), viii) Administration and Management (leadership, human resource, and financial management skill)

Snell & Bohlander (2012) use many words to describe how important people are for organizations. The terms human resources, human capital, intellectual assets, and talent management imply that it is driving the performance of their organizations (along with other resources such as money, materials and information). Successful libraries are adept at bringing together different people to achieve a common purpose.

According to Sharma (2014), leadership skills are:
i) Technical Skills ii) Conceptual Skill iii) Interpersonal Skills iv) Emotional Intelligence v) Social Intelligence vi) Constant Learning

Thanuskodi (2015) summarized competencies required by library and information professionals:
(1) Philosophical competencies; (2) Technological competencies; (3) Educational/professional and personal competencies; (4) Customer service competencies; (5) Administration and leadership related competencies; and (6) Information literacy skills

Yang et al. (2016) reviewed librarian job advertisements (2009-2014). They found out some skills required for the librarian, which are "Communication skills, project management, work collaboratively, leadership, cataloguing, Analytical, and problem-solving skills, etc." They used a content analysis method for their research.

Biola (2017) Leadership as "a process whereby an individual influences a group of individuals to achieve a common goal" (in-text citation). Northouse (2001) determined four general themes in the way leadership works: (1) leadership is a process; (2) leadership involves influence; (3) leadership occurs in a group context; (4) leadership involves goal attainment.

As Whetten and Cameron (2017) stated, management skills consist of identifiable sets of actions that an individual performs and lead to specific outcomes. Ranganathan (1931) "The library is a growing organism" so as a librarian updates himself/herself, by the way, all the managerial aspects of library resources grow.

Ali and Richardson (2018) discussed the vital information of literacy skills in workplace system of information usage in library software, online information resources accessing knowledge various databases, information searching techniques such as how to retrieve information from search engines, information research support usage and awareness of plagiarism software, information literacy competencies Web 2.0 skills, generic competencies and personal attributes.

Cherinet (2018) in his survey, states that the skills of librarians are hard skills (technical knowledge), soft skills, trans-literacy, leadership Skills, analytics skills, civic literacy, research skills, twenty-first-century skills, life and career skills, survival skills, global awareness, political skills, cultural Intelligence, collaborations skill, innovation skills, business skills, marketing skills, negotiation skills, design skill, learning skills, discipline, passions, professionalism, etc.

Oza and Mehta (2018), concentrate on ICT skills and competencies needed for corporate information professionals. In their survey, they found that LIS professionals should have their own professional and personal skills. Librarians may play different roles like information enabler, information architect, knowledge engineer, knowledge creator, database & network creator, web designer, online publisher, decision support specialist.

Malik and Ameen (2021), their study finds that competencies needed for LIS professionals are subject knowledge & skills, i.e., subject in both theoretical and practical aspects. IT knowledge skills are also essential in-case of teaching and delivering IT-related course content. In addition, they require instructional skills for reaching out to the user's needs. Generic skills of LIS Professionals are management, leadership, and communication. The way of speaking is more important in communication. Along with this, self-learning plays a vital role.

### 3. Methodology

The author adopted the quantitative and qualitative methods; it is a survey method and a descriptive study. We collected data through the help of Google Form. Investigators collect the Library and Information Science Professionals Email IDs through their Institutes of national importance and central universities official websites and request mail and reminder mails were sent to fill the questionnaire. It's a random sampling approach. The SPSS software tool used for data analysis. Frequency, percentage, mean and standard deviation were used in a descriptive analysis of the data. The questionnaire was divided into two sections: i) demographics and ii) advanced level of literacy skills. A questionnaire consisting of 12 questions, demographic questions five and literacy-based questions seven. Likert scale was used for calculation mean and standard deviation. We limited the study to the Institute of National Importance in India and Central Universities in India.

### 4. Data Analysis and Findings

**Table.1 Demographics characteristics of Respondents (N= 223)**

| Type | Categories | n (%) |
|---|---|---|
| Gender | Female | 66 (29.6%) |
| | Male | 157 (70.4%) |
| Institute/ University | Institute of National Importance | 156 (70%) |
| | Central Universities | 67 (30%) |
| Educational Qualifications | Undergraduate | 3 (1.3%) |
| | Postgraduate | 143 (64.1%) |
| | M.Phil. | 16 (7.2%) |
| | PhD | 61 (27.4%) |
| Designation | Librarian | 21 (9.4%) |
| | Deputy Librarian | 13 (5.8%) |
| | Assistant Librarian | 70 (31.4%) |
| | Library Assistant | 38 (17%) |
| | Professional Assistant | 36 (16.2%) |
| | Technical Assistant | 18 (8.1%) |
| | Library Trainee | 27 (12.1%) |
| Experience | Less than 1 year | 12 (5.4%) |
| | 1-5 years | 59 (26.4%) |

|  | 6-10 years | 49 (22%) |
|---|---|---|
|  | 11-20 years | 60 (26.9%) |
|  | Over 20 years | 43 (19.3%) |

Table 1 revealed the gender-wise respondents. 70.4 % of the responders are male, and female respondents are 29.6%. The above table shows that 156 (70%) respondents from institutes of National Importance (Such as IITs, IIITS, IIMs, NITs, IISER and Others, 67 (30%) respondents are Central Universities in India. The highest educational qualifications of respondents. 1.3% are Graduates, 64.1% Postgraduates, 7.2% are Master of Philosophy, and 27.4% are Doctorates. It shows us where the highest respondents completed their Postgraduates 143 (64.1%), the least is Graduates 3 (1.3%). The above table result providing the information is in Indian Library professionals most of them minimum possessed with Postgraduate. According to the designation-wise distribution of the respondents, it's clear that most of the respondents are Assistant Librarian 70 (31.4%) and the least respondents are Technical Assistant 18 (8.1%). It can be seen from the result given above Table that the highest respondents are 26.9% they are 11 to 20 years' experience, followed by 26.4% respondents are 1 to 5 years of experience and the least respondents 5.4% they are less than one year experience.

### Table 2. Data Skills and Their Awareness

| Level of Awareness in Data Skills | Not at all aware | Slightly aware | Somewhat aware | Moderately aware | Extremely aware | Mean | SD |
|---|---|---|---|---|---|---|---|
| Metadata and Documentation | 10 (4.5%) | 6 (2.7%) | 20 (9%) | 96 (43%) | 91 (40.8%) | 4.13 | 0.99 |
| Data Visualization | 16 (7.2%) | 13 (5.8%) | 46 (20.6%) | 91 (40.8%) | 57 (25.6%) | 3.71 | 1.12 |
| Data Use and Reuse | 4 (1.8%) | 12 (5.4%) | 34 (15.2%) | 102 (45.7%) | 71 (31.9%) | 4 | 0.92 |
| Data Storage and Preservation | 7 (3.1%) | 8 (3.6%) | 24 (10.8%) | 82 (36.8%) | 102 (45.7%) | 4.18 | 0.97 |
| Data Discovery | 9 (4%) | 22 (9.9%) | 41 (18.4%) | 80 (35.9%) | 71 (31.8%) | 3.81 | 1.1 |
| Data Literacy | 6 (2.7%) | 23 (10.3%) | 39 (17.5%) | 81 (36.3%) | 74 (33.2%) | 3.86 | 1.06 |
| Data Structures and Standards | 11 (5%) | 41 (18.4%) | 45 (20.2%) | 76 (34%) | 50 (22.4%) | 3.5 | 1.16 |
| Data Publication & Citation | 9 (4%) | 25 (11.2%) | 43 (19.3%) | 58 (26%) | 88 (39.5%) | 3.85 | 1.17 |
| Linked Data | 16 (7.2%) | 38 (17%) | 41 (18.4%) | 71 (31.8%) | 57 (25.6%) | 3.51 | 1.23 |
| Data Repository | 9 (4%) | 17 (7.6%) | 25 (11.2%) | 78 (35%) | 94 (42.2%) | 4.03 | 1.09 |

Scale Used: 1 Not at all aware; 2 Slightly aware; 3 Somewhat aware; 4 Moderately aware; 5 Extremely aware; M: Mean SD: Standard Deviation

Table 2 shows the data skills of the respondents. Most of the respondents 43% moderately aware of metadata and documentation skills, 40.8% respondents moderately aware of data visualization, 45.7% respondents moderately aware of data use and reuse, 45.7% respondents extremely aware of data storage and preservation, 35.9% respondents moderately aware of data discovery, 36.3% respondents moderately aware on data literacy, 34% of respondents moderately aware on data structure and standards, 39.5% of respondents extremely aware on data publication and citation, 31.8% of respondents moderately aware on linked data, and 35% of respondents moderately aware on the data repository.

**Table 3. Necessary Skills for Library Professionals**

| Important skills for Library Professionals | Not at all important | Slightly important | Important | Fairly important | Very important | Mean | SD |
|---|---|---|---|---|---|---|---|
| Digital Literacy skills | 6 (2.7%) | 2 (0.9%) | 33 (14.8%) | 35 (15.7%) | 147 (65.9%) | 4.41 | 0.95 |
| Soft skills | 6 (2.7%) | 8 (3.6%) | 11 (4.9%) | 42 (18.8%) | 156 (70%) | 4.49 | 0.94 |
| Organizational skills | 6 (2.7%) | 8 (3.6%) | 22 (9.9%) | 36 (16.1%) | 151 (67.7%) | 4.42 | 0.98 |
| Negotiation skills | 0 (0%) | 8 (3.6%) | 41 (18.4%) | 74 (33.2%) | 100 (44.9%) | 4.19 | 0.85 |
| Marketing skills | 0 (0%) | 10 (4.5%) | 47 (21.1%) | 61 (27.4%) | 105 (47%) | 4.17 | 0.91 |
| Leadership skills | 0 (0%) | 5 (2.2%) | 23 (10.3%) | 39 (17.5%) | 156 (70%) | 4.55 | 0.76 |
| English language skills | 2 (0.9%) | 11 (5%) | 43 (19.3%) | 50 (22.4%) | 117 (52.4%) | 4.2 | 0.97 |

Scale Used: 1 Not at all important; 2 Slightly important; 3 Important; 4 Fairly important; 5 Very important; M: Mean SD: Standard Deviation

## Figure 1. Important Skills for Library Professionals

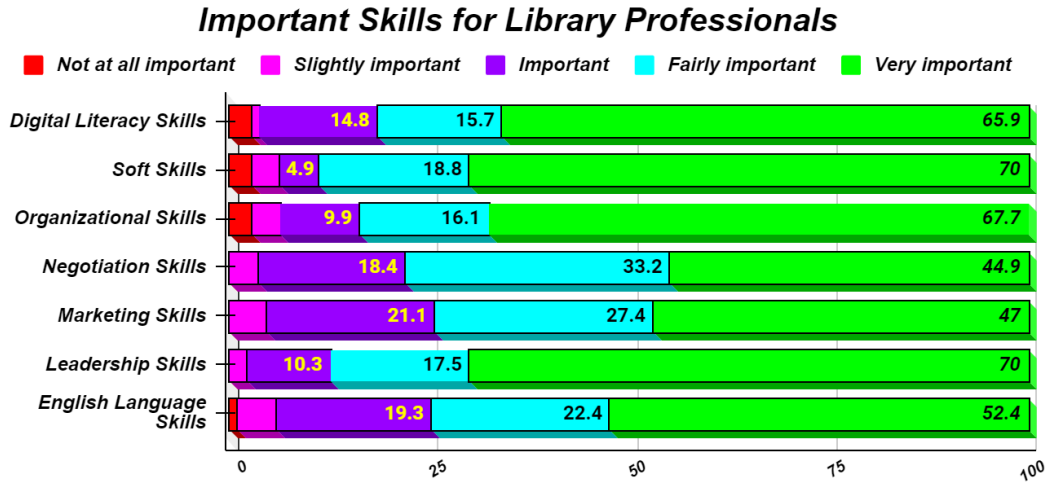

Table 3 and figure 1 represent the important skills of library professionals. The highest respondents 65.9% says digital literacy skills are very important, 70% of respondents say soft skills are very important, 67.7 of respondents says organization skill is very important, 44.9% of respondents say negotiation skills are very important, 47% respondents say that marketing skill is very important, 70% of respondents say leadership skills are very important, 52.4% of respondents say English language skills are very important.

### Table 4. ICT Competencies of Library Staff

| ICT Competencies of Library Staff | Not at all familiar | Slightly familiar | Somewhat familiar | Moderately familiar | Extremely familiar | Mean | SD |
|---|---|---|---|---|---|---|---|
| Automation and digitization skills | 0 (0%) | 5 (2.3%) | 7 (3.1%) | 54 (24.2%) | 157 (70.4%) | 4.62 | 0.65 |
| Presentation skills | 0 (0%) | 10 (4.5%) | 8 (3.6%) | 81 (36.3%) | 124 (55.6%) | 4.43 | 0.76 |
| Web 2.0/Library 2.0 skills | 7 (3%) | 16 (7.2%) | 26 (11.7%) | 80 (35.9%) | 94 (42.2%) | 4.06 | 1.05 |
| Technical skills | 5 (2.2%) | 23 (10.3%) | 51 (22.8%) | 65 (29.2%) | 79 (35.5%) | 3.85 | 1.08 |
| Use of electronic resources/Multimedia | 0 (0%) | 5 (2.2%) | 6 (2.7%) | 53 (23.8%) | 159 (71.3%) | 4.64 | 0.64 |

Scale Used: 1 Not at all familiar; 2 Slightly familiar; 3 Somewhat familiar; 4 Moderately familiar; 5 Extremely familiar; M: Mean SD: Standard Deviation

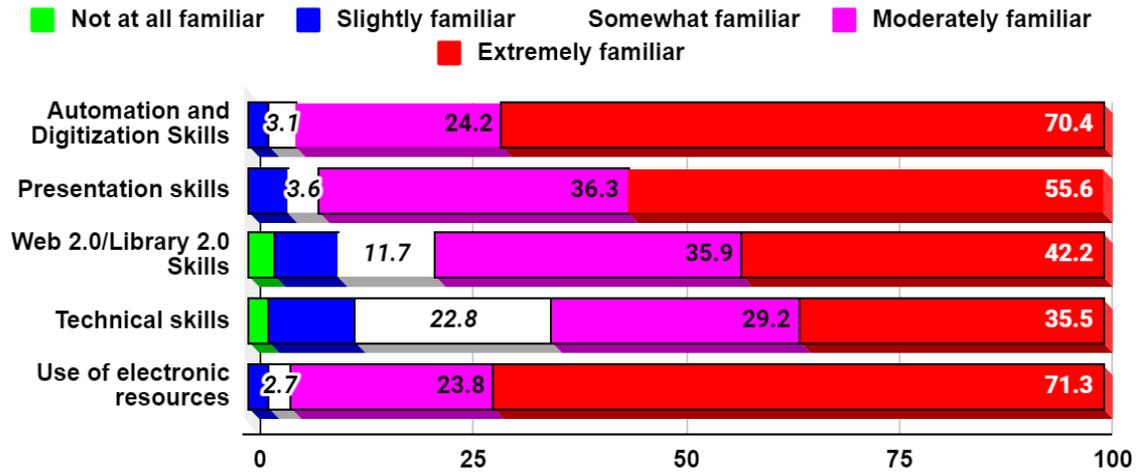

**Figure 2. ICT Competencies of Library Staff**

Table 4 and figure 2 revealed the ICT competencies of library staff. Most of the respondents 70.4% extremely familiar with automation and digitization skills, 55.6% of respondents extremely familiar with presentation skills, 42.2% of respondents extremely familiar with web2.0/library skills, 35.5% of respondents extremely familiar with technical skills, 71.3% respondents extremely familiar with the use of electronic resources.

**Table 5. Digital Literacy Skills and Their Familiarity**

| Digital Literacy Skills Level of Familiarity | Not at all familiar | Slightly familiar | Somewhat familiar | Moderately familiar | Extremely familiar | Mean | SD |
|---|---|---|---|---|---|---|---|
| Artificial Intelligence | 9 (4%) | 53 (23.8%) | 60 (26.9%) | 83 (37.2%) | 18 (8.1%) | 3.21 | 1.02 |
| Big Data | 15 (6.7%) | 54 (24.2%) | 54 (24.2%) | 76 (34.1%) | 24 (10.8%) | 3.17 | 1.11 |
| Cyber Security | 7 (3.1%) | 54 (24.2%) | 39 (17.5%) | 74 (33.2%) | 49 (22%) | 3.41 | 1.11 |
| Internet of Thing (IoT) | 15 (6.7%) | 33 (14.8%) | 52 (23.3%) | 86 (38.6%) | 37 (16.6%) | 3.43 | 1.13 |
| Virtual Reality | 17 (7.6%) | 37 (16.6%) | 53 (23.7%) | 77 (34.6%) | 39 (17.5%) | 3.37 | 1.17 |
| Cloud computing | 14 (6.3%) | 39 (17.5%) | 32 (14.3%) | 74 (33.2%) | 64 (28.7%) | 3.6 | 1.24 |

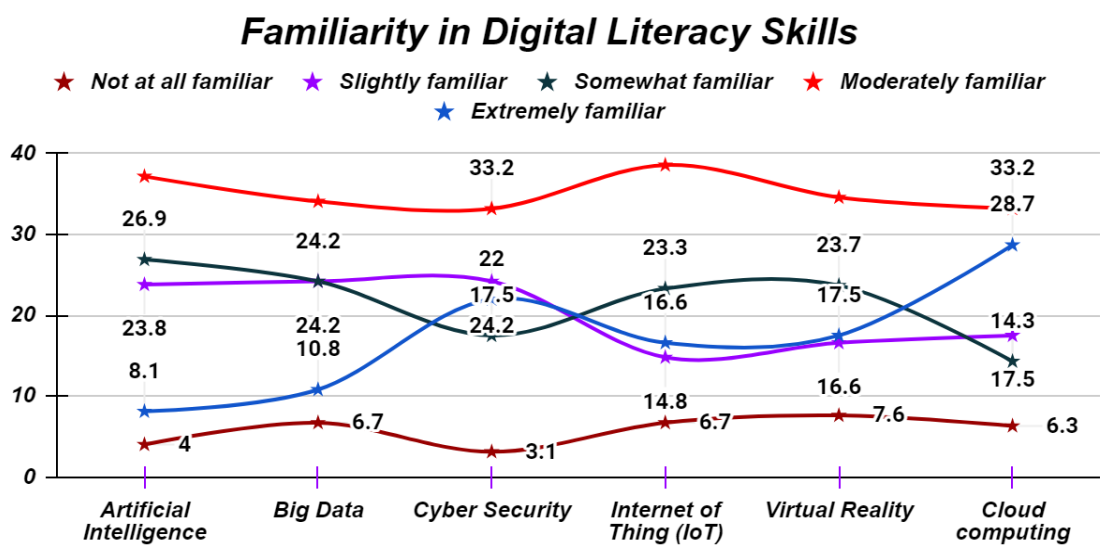

**Figure 3. Digital Literacy Skills and Their Familiarity**

Table 5 and figure 3 show that present digital literacy skills and their level of familiarity. Most of the respondents, 37.2% moderately familiar with artificial intelligence, 34.1% respondents moderately familiar with big data, 33.2% of respondents moderately familiar with cybersecurity, 38.6% of respondents moderately familiar on Internet of Things, 34.6% of respondents moderately familiar with virtual reality and 33.2% of respondents moderately familiar on cloud computing.

**Table 6. Working Confidence Level of Integrated Library Management System, Institutional Repository, Content Management System Software, Reference Management Software/Citation Management Tools**

| Confidence in performing ILMS, IR, CMS, and RMS | I don't think I could do that | I could probably manage that | Fair/OK | Confident enough to show/tell others | This does not apply in my work environment |
|---|---|---|---|---|---|
| Koha | 23 (10.3%) | 2 (0.9%) | 28 (12.6%) | 139 (62.3%) | 31 (13.9%) |
| Evergreen | 34 (15.2%) | 69 (31%) | 31 (13.9%) | 16 (7.2%) | 73 (32.7%) |
| NewGenLib | 31 (13.9%) | 66 (29.6%) | 39 (17.5%) | 27 (12.1%) | 60 (32.7%) |
| Libsys | 13 (5.8%) | 33 (14.8%) | 29 (13%) | 92 (41.3%) | 56 (25.1%) |
| SOUL | 12 (5.4%) | 52 (23.3%) | 52 (23.3%) | 53 (23.8%) | 54 (24.2%) |
| Dspace | 3 (1.3%) | 23 (10.3%) | 50 (22.4%) | 115 (51.6%) | 32 (14.4%) |
| GreenStone | 23 (10.3%) | 49 (22%) | 59 (26.5%) | 42 (18.8%) | 50 (22.4%) |
| EPrints | 27 (12.1%) | 56 (25.1%) | 41 (18.4%) | 47 (21.1%) | 52 (23.3%) |
| Fedora | 40 (17.9%) | 65 (29.1%) | 37 (16.6%) | 19 (8.6%) | 62 (27.8%) |
| WordPress | 18 (8.1%) | 52 (23.4%) | 42 (18.8%) | 75 (33.6%) | 36 (16.1%) |
| Joomla | 22 (9.9%) | 55 (24.6%) | 66 (29.5%) | 50 (22.5%) | 30 (13.5%) |
| Drupal | 25 (11.2%) | 46 (20.6%) | 64 (28.7%) | 56 (25.1%) | 32 (14.4%) |
| Mendeley | 13 (5.8%) | 42 (18.9%) | 41 (18.3%) | 105 (47.1%) | 22 (9.9%) |
| EndNote | 15 (6.7%) | 49 (22%) | 59 (26.5%) | 74 (33.2%) | 26 (11.6%) |
| Zotero | 16 (7.2%) | 40 (18%) | 61 (27.3%) | 88 (39.4%) | 18 (8.1%) |

As shown in table 6, the working confidence level of the library and related software and tools. Confident enough to show/tell others the majority of the respondents 62.3% Koha is ILMS, 51.1% respondents in Dspace is IR, 33.6% of respondents WordPress is CMS, and 47.1% of respondents in Mendeley is RMS.

**Table 7. Way of gained ICT Skills**

| Acquired ICT Skills | Frequency | Percentage |
|---|---|---|
| Computer/ICT Training Centre/Cyber café | 138 | 61.9 |
| Library Schools | 83 | 37.2 |
| On the-job Training | 151 | 67.7 |
| Personal Practice | 181 | 81.2 |
| Additional University Qualification in Computer | 83 | 37.2 |
| Workshop/Seminars/Conferences | 169 | 75.8 |

Table 7 shows that library professionals gained ICT skills. The majority of the responded 81.2% acquired by personal practice, followed by Workshop/Seminar/Conference responders was 75.8%.

**Table 8. Factors Contributing to Inefficient ICT Competencies for Effective Job Performances and Barriers to job performance in the library**

| Factors Contributing to Inefficient ICT Competencies | Strongly Agree | Agree | Neutral | Disagree | Strongly Disagree | Mean | SD |
|---|---|---|---|---|---|---|---|
| Inadequate ICT facilities | 83 (37.2%) | 70 (31.4%) | 36 (16.2%) | 27 (12.1%) | 7 (3.1%) | 3.87 | 1.13 |
| Inaccessibility to ICT facilities | 55 (24.7%) | 99 (44.4%) | 31 (13.9%) | 32 (14.3%) | 6 (2.7%) | 3.73 | 1.06 |
| Negative attitude of library staff | 50 (22.4%) | 64 (28.7%) | 47 (21.1%) | 49 (22%) | 13 (5.8%) | 3.39 | 1.21 |
| Age | 20 (9%) | 64 (28.7%) | 71 (31.8%) | 49 (22%) | 19 (8.5%) | 3.07 | 1.09 |
| Lack of acknowledgement for work done | 30 (13.4%) | 104 (46.6%) | 63 (28.3%) | 21 (9.4%) | 5 (2.3%) | 3.59 | 0.91 |
| Work overload due to shortage of staff | 54 (24.2%) | 107 48(%) | 34 (15.3%) | 18 (8.1%) | 10 (4.4%) | 3.79 | 1.03 |
| Lack of contingent rewards and wages | 46 (20.6%) | 86 (38.6%) | 61 (27.3%) | 18 (8.1%) | 12 (5.4%) | 3.6 | 1.06 |
| Lack of commitment to career and capacity development | 57 (25.5%) | 69 (31%) | 63 (28.3%) | 27 (12.1%) | 7 (3.1%) | 3.63 | 1.08 |
| Lack of recognition and status | 58 (26%) | 71 (31.8%) | 66 (29.6%) | 23 (10.3%) | 5 (2.3%) | 3.69 | 1.03 |
| Lack of training and retraining programmes | 78 (35%) | 92 (41.3%) | 29 (13%) | 21 (9.4%) | 3 (1.3%) | 3.99 | 0.98 |

Scale Used: 1 Strongly disagree; 2 Disagree; 3 Neutral; 4 Agree; 5 Strongly agree; M: Mean SD: Standard Deviation

Above Table 8 and below figure 4 evidences of factors contributing to inefficient ICT competencies for effective job performances and barriers to job performance in the library. The majority of the respondents agreed with 48.1% work overload because of the shortage of staff, followed by a lack of acknowledgement of 46.6% of respondents.

**Figure 4. Factors of contributing to inefficient ICT competencies**

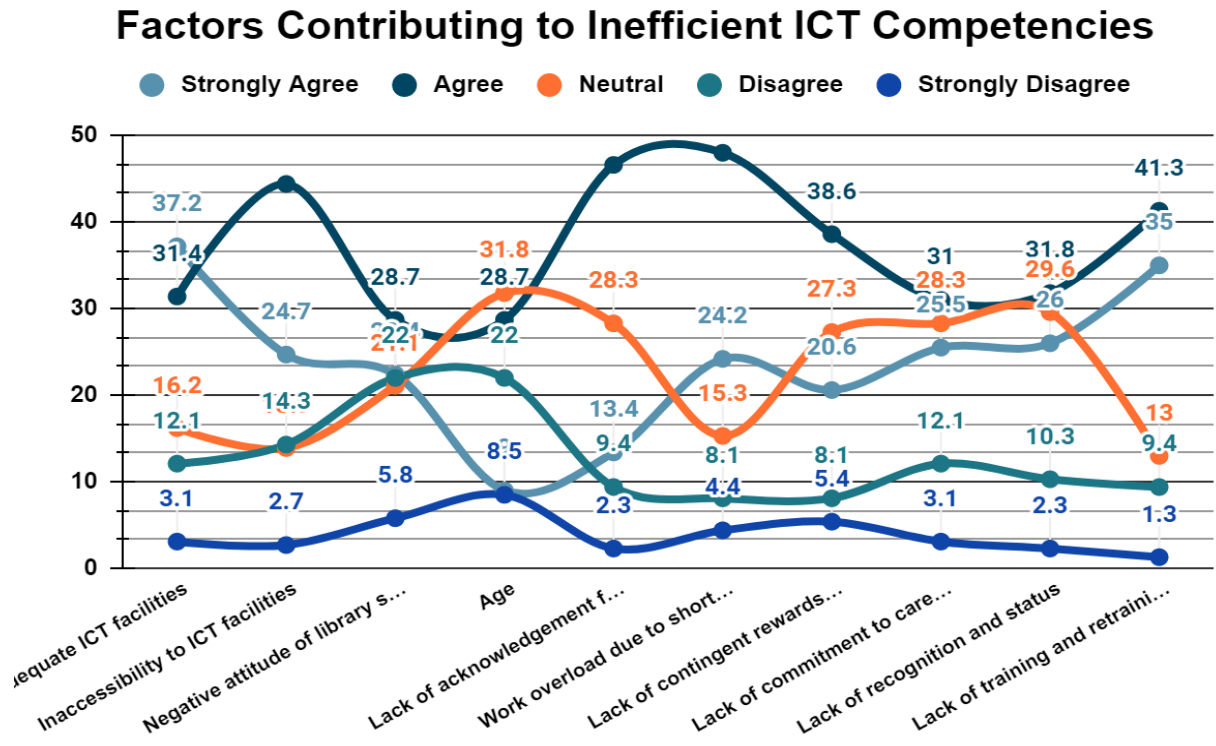

## 5. Conclusion

The result found that Indian library professionals are updated professionals in both foundational competencies and functional competencies. It is commonly known that while working by learning with personal practice as the majority of the librarians learned with their self-interest and practice. Library professionals are hyperactive in attending and conducting conferences/webinars/seminars/workshop/orientation programmes and so on; they are also growing themself as well as helping to grow the user community. The library is a non-profit organization but library professional's only expectation is acknowledging and appreciation for their work, not more than that. The users, as well as colleagues, have to encourage them in their work. They ran some institutes with a staff shortage, which adds a burden to the library professionals in the working environment.

**References**
1. ALA. (2010, August 4). *ALA's Core Competence of Librarianship (Old Number 40.3)* http://www.ala.org/aboutala/governance/policymanual/updatedpolicymanual/section2/40 corevalues
2. Ali, M. Y., & Richardson, J. (2018). Workplace Information Literacy Skills: Library Professionals' Competency at University Libraries in Karachi, Pakistan. *Information and Learning Science*, *119*(7/8), 469–482. https://doi.org/10.1108/ILS-10-2017-0107
3. Atanda, A. D., Owolabi, K. A., & Ugbala, C. P. (2021). Professional competence and attitudes of library personnel towards digital services in selected university libraries in Nigeria. *Digital Library Perspectives*, *ahead-of-print*(ahead-of-print). https://doi.org/10.1108/DLP-08-2020-0076
4. Biola, A. (2017, August 9). *Leadership: A Practical Meaning*. Medium. https://medium.com/@adimulabiola/leadership-a-practical-meaning-8ee669f4887a
5. Cherinet, Y. M. (2018). Blended Skills and Future Roles of Librarians. *Library Management*, *39*(1/2), 93–105. https://doi.org/10.1108/LM-02-2017-0015
6. Dierdorff, E. C. (2020, January 29). Time Management Is About More Than Life Hacks. *Harvard Business Review*. https://hbr.org/2020/01/time-management-is-about-more-than-life-hacks
7. Huang, Y., Cox, A. M., & Sbaffi, L. (2021). Research Data Management Policy and Practice in Chinese University Libraries. *Journal of the Association for Information Science and Technology*, *72*(4), 493–506. https://doi.org/10.1002/asi.24413
8. Kwan, D., & Shen, L. (2015). Senior Librarians' Perceptions of Successful Leadership Skills. In *Advances in Library Administration and Organization* (Vol. 33, pp. 89–134). Emerald Group Publishing Limited. https://doi.org/10.1108/S0732-067120150000033003
9. Malik, A., & Ameen, K. (2021). Needed Competencies for Library and Information Faculty Members in Pakistan. *Library Philosophy and Practice (e-Journal)*. https://digitalcommons.unl.edu/libphilprac/5186
10. Northouse, P. G. (2001). *Leadership: Theory and Practice*. SAGE Publications, Inc.
11. Northouse, P. G. (2012). *Leadership: Theory and Practice* (Sixth edition). SAGE South Asia.
12. Oza, D. N., & Mehta, M. (2018). A study of ICT Skills and Competencies Essential for Corporate LIS Professionals. *International Journal of Research and Analytical Reviews(IJRAR)*, *5*(4), 8.
13. Ranganathan, S. R. (1931). *The Five Laws of Library Science*. Madras Library Association (Madras, India) and Edward Goldston (London, UK). https://repository.arizona.edu/handle/10150/105454
14. S. Thanuskodi. (2015). *Professional Competencies and Skills for Library and Information Professionals* (pp. 1–9) [Working Paper]. The Congress of Southeast Asian Librarians (CONSAL). https://myrepositori.pnm.gov.my/handle/123456789/4188
15. Saibakumo, W. (2021). Awareness and Acceptance of Emerging Technologies for Extended Information Service Delivery in Academic Libraries in Nigeria. *Library Philosophy and Practice (e-Journal)*, 1–11.
16. Sharma, D. (2014). *Leadership Lessons from the Military*. https://doi.org/10.4135/9789351508045
17. Snell, S., & Bohlander, G. W. (2012). *Managing Human Resources* (16th edition). South-Western College Publishing.
18. Ugwu, C., & Ezema, I. (2010). *Competencies for Successful Knowledge Management Applications in Nigerian Academic Libraries*. *2*.
19. Whetten, D. A., & Cameron, K. S. (2017). *Developing Management Skills* (Eighth edition). Pearson Education.
20. Yang, Q., Zhang, X., Du, X., Bielefield, A., & Liu, Y. Q. (2016). Current Market Demand for Core Competencies of Librarianship—A Text Mining Study of American Library Association's Advertisements from 2009 through 2014. *Applied Sciences*, *6*(2), 48. https://doi.org/10.3390/app6020048